# Performance of Silicon Immersed Gratings: Measurement, Analysis & Modelling

Michiel Rodenhuis*[a], Paul J. J. Tol[a], Tony H. M. Coppens[a], Phillip P. Laubert[a], Aaldert H. van Amerongen[a]

[a]SRON Netherlands Institute for Space Research, Sorbonnelaan 2, 3584CA Utrecht, The Netherlands;

**ABSTRACT**

The use of Immersed Gratings offers advantages for both space- and ground-based spectrographs. As diffraction takes place inside the high-index medium, the optical path difference and angular dispersion are boosted proportionally, thereby allowing a smaller grating area and a smaller spectrometer size. Short-wave infrared (SWIR) spectroscopy is used in space-based monitoring of greenhouse and pollution gases in the Earth atmosphere. On the extremely large telescopes currently under development, mid-infrared high-resolution spectrographs will, among other things, be used to characterize exo-planet atmospheres.

At infrared wavelengths, Silicon is transparent. This means that production methods used in the semiconductor industry can be applied to the fabrication of immersed gratings. Using such methods, we have designed and built immersed gratings for both space- and ground-based instruments, examples being the TROPOMI instrument for the European Space Agency Sentinel-5 precursor mission, Sentinel-5 (ESA) and the METIS (Mid-infrared E-ELT Imager and Spectrograph) instrument for the European Extremely Large Telescope.

Three key parameters govern the performance of such gratings: The efficiency, the level of scattered light and the wavefront error induced. In this paper we describe how we can optimize these parameters during the design and manufacturing phase. We focus on the tools and methods used to measure the actual performance realized and present the results.

In this paper, the bread-board model (BBM) immersed grating developed for the SWIR-1 channel of Sentinel-5 is used to illustrate this process. Stringent requirements were specified for this grating for the three performance criteria. We will show that –with some margin– the performance requirements have all been met.

## 1. INTRODUCTION

As Silicon is transparent to light with wavelengths in excess of 1.2 μm and has a high index of refraction of 3.4 it is an attractive material for diffractive optics in the infrared. A particular example of this is its use in immersed gratings. When illuminating the grating pattern from the inside, the dispersive power is boosted proportionally by the reduced wavelength in the high index of refraction medium. When used in a spectrograph, the immersed grating thus has the effect of drastically increasing the spectral resolution that can be achieved for a spectrograph of a given size (see Figure 1). Inversely, a spectrograph with a certain required resolution can be made much smaller, a prospect that is particularly attractive for use in space-borne applications.

Silicon offers a number of additional advantages: Its widespread use in the semiconductor industry means that a large body of knowledge and techniques on microfabrication, in particular using photolithography, is available. These techniques can readily be used in the fabrication of the grating pattern. After photolithographic masking, the anisotropy of monocrystaline Silicon can be used to etch a blazed grating groove resulting in an optimized grating efficiency. This etching yields a further advantage: The resulting groove surfaces are very smooth, allowing gratings with very low induced wavefront aberration to be manufactured.

At SRON we have adopted this technology to produce a number of immersed gratings for different projects. The first grating produced was for the short-wave infrared channel of the TROPOMI instrument, soon to be launched on the ESA

*m.rodenhuis@sron.nl; www.sron.nl

Sentinel-5 precursor mission. For this grating, the groove pattern was etched in a cylindrical block ('puck') of Silicon out of which the prism shape was then cut. A Technology Readiness Level (TRL) of 8 was achieved for this method of immersed grating production[1]. For later gratings, an improved process whereby the grating pattern is etched into a semiconductor industry-standard silicon wafer and then bonded to the prism has been adopted. The advantage of this technique is that many standard tools and processes are available for photolithography on these wafers. Also, separating the manufacturing of the grating wafer from the prism allows a number of grating wafers to be produced and characterized after which the best can be chosen for bonding with the prism. Using this method, two immersed gratings have been produced for the bread-board model (BBM) of the short-wave infrared 1 (SWIR-1) channel of the spectrograph for the full Sentinel-5 mission. TRL of 5 was achieved in this pre-development project. A very large (150x126x101 mm) demonstrator model immersed grating has also been produced for the ground-based METIS infrared spectrograph that is under development for the future European Extremely Large Telescope (E-ELT)[2,3]. The new manufacturing method, using a 150-mm wafer, was essential to make the production of such a large grating possible. We have recently started the production of a large number of immersed gratings for both the SWIR-1 and SWIR-3 channels of the Sentinel-5 mission.

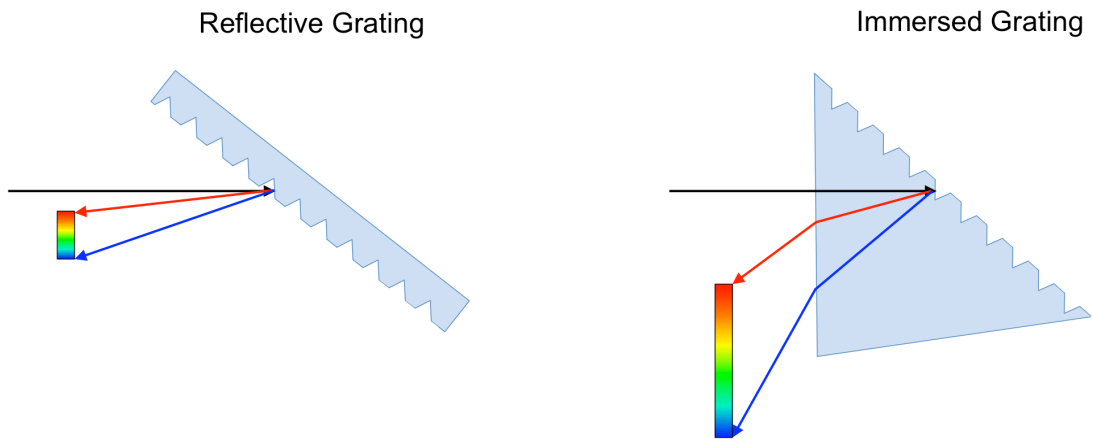

**Figure 1: Schematic representation of the advantage of the high index of refraction immersed grating over a traditional reflective grating.**

This paper describes the design, manufacture and performance characterization of immersed gratings by SRON, focusing on the SWIR-1 BBM for Sentinel-5. The key performance parameters and the design methodology implemented to optimize them are described in the next section. In section 3 we describe the manufacturing process. In section 4 we detail the methods used to inspect and characterize the (lithographic) quality of the fabricated grating wafers. Section 4 then discusses the performance testing and the results obtained, after which we present the conclusions of this paper.

## 2. IMMERSED GRATING DESIGN AND KEY PERFORMANCE PARAMETERS

In the design of an immersed grating, the overarching goal is to achieve a certain amount of angular dispersion for a given wavelength range. As with a classical grating, this is controlled through the grating period and the dispersion order in which it is operated. Beyond this primary requirement, the three key performance parameters that must be optimized are:

- The grating efficiency, specified separately for two orthogonal linear polarization states
- The WaveFront Error (WFE) imparted on the incoming beam by the grating
- The amount of unwanted light scattered by the grating, in particular in the outgoing dispersed beam.

The efficiency is primarily optimized with the blaze angle and the so-called dam width, the width of the groove-less part of the period. These and other parameters of the grating geometry are shown in Figure 2. In the fabrication of the grooves we make use of the anisotropy of the silicon crystal lattice. Using a potassium hydroxide solution, the etching speed perpendicular to the {1,1,1} crystal plane is about hundred times slower than parallel to it, with the exact etching speed difference depending on the concentration used and environmental factors such as temperature. The desired blaze angle is achieved through controlling the so-called off-cut angle, the angle between the {1,0,0} crystal plane and the grating wafer surface. This angle is specified when ordering the wafers for the grating manufacture. We use the grating simulation package *PCGrate*[4], embedded in our own simulation software[5] to optimize the blaze angle and dam width for a specific design requirement.

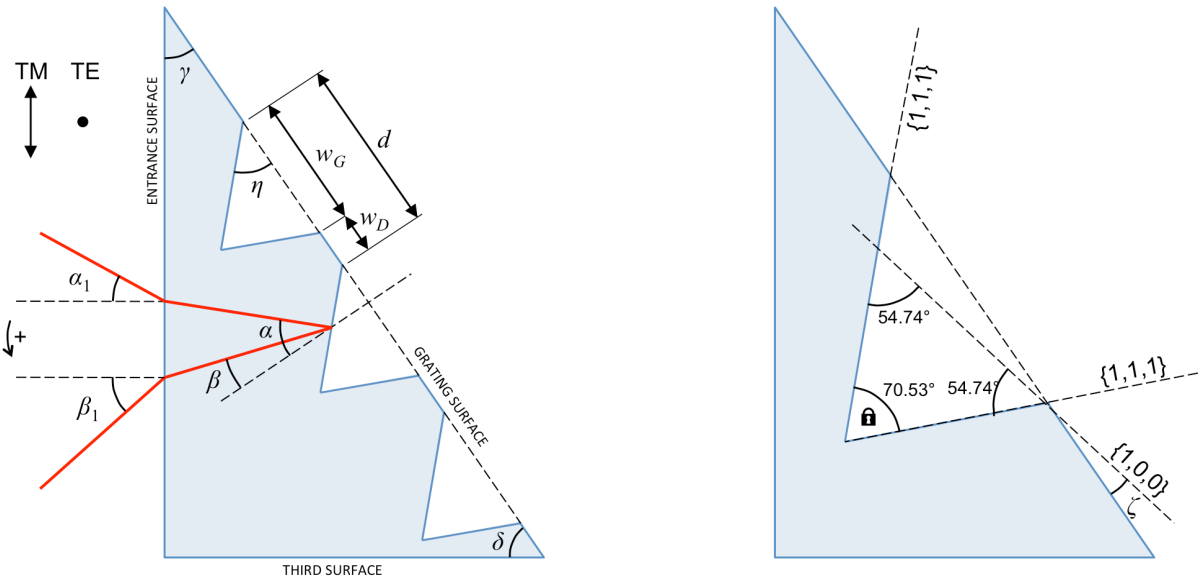

**Figure 2: Geometry of the blazed grating pattern etched in monocrystalline silicon. The right figure shows how the blazed grooves are defined by the orientation of the principal planes {1,1,1} and {1,0,0} in the silicon crystal lattice.**

In Figure 2 all angles are defined in the dispersion plane, i.e. the plane perpendicular to the grating grooves. Angle $\gamma$ is the top angle of the prism, between the input surface and the grating surface. Angle $\delta$ is the angle between the grating surface and the third surface. The angle of incidence on the IG is $\alpha_1$ and on the grating surface $\alpha$. The diffraction angle is $\beta$ and the output angle is $\beta_1$. The V-shaped grooves are separated by a flat 'dam' part of width $w_D$, which makes up less than one third of the grating period $d$. The grooves are asymmetric, with blaze angle $\eta$ between the grating surface and the groove facet that is used for diffraction. The angles of incidence and diffraction are centred on the blaze angle, such that the grating operates near the optimal Littrow configuration. TM waves have the E-field component pointing in the plane of the figure; TE waves have an E-field pointing out of the figure. The angle between the {1,0,0} crystal axis of the silicon and the grating surface is $\zeta$, positive in the direction of the {1,1,1} axis. This is known as the off-cut angle.

The WFE of the immersed grating is determined by the polishing quality of the prism, the flatness of the wafer and the flatness of the etched grating surface. When expressed as an rms deviation, the individual WFE components add in quadrature as expressed mathematically in eq. 1. Here $W_{IG}$ is the total (rms) WFE of the immersed grating, $n$ the index of refraction, $W_{PE}$ and $W_{PG}$ the rms WFE of the prism entrance and grating surfaces, $W_{TTV}$ the (rms) total thickness variation of the grating wafer and $W_G$ the WFE introduced by the grating pattern and groove surfaces. In the expression, the WFE quantities are all as measured externally (not in immersion) and $\alpha$, $\beta$, $\alpha_1$ and $\beta_1$ are assumed to be small (Littrow configuration). Also, $W_{PG}$, $W_{TTV}$, and $W_G$ are assumed to be correlated for the incoming and outgoing beam while $W_{PE}$ is uncorrelated. $W_G$ is measured in the diffraction order used and therefore does not require multiplication by $\cos \eta$.

$$W_{IG} = \sqrt{\left[\sqrt{2}(n-1)W_{P_E}\right]^2 + \left[2nW_{P_G}\cos\eta\right]^2 + \left[2nW_{TTV}\cos\eta\right]^2 + \left[2nW_G\right]^2} \tag{1}$$

The component WG depends heavily on the straightness of the grooves and the uniformity of the groove period, so on the precision of the lithographic process. This will be discussed further in Section 4. For the polishing flatness of the silicon prism, which we source externally, we typically require a WFE of 10 nm rms.

One of the attractions of etching the grating pattern in monocrystalline silicon is the extremely smooth groove surfaces that can be achieved. This results in relatively low levels of randomly scattered light. The unwanted reflected (stray) light is minimized by optimizing the optical (antireflection coatings) and designing the prism to avoid light from non-nominal internal reflections ending up in the dispersed beam. In addition, surface roughness and dust contamination of the prism entrance surface must be kept as low as possible. We apply an antireflection coating to the entrance surface, a reflecting coating to the grating surface and an absorbing coating to the third surface. The simulation and design suite that we have developed in the Mathematica framework is used to perform ray-tracing calculations to optimize the prism design. Results from such simulations for the Sentinel-5 SWIR-1 prism are presented in Figure 3.

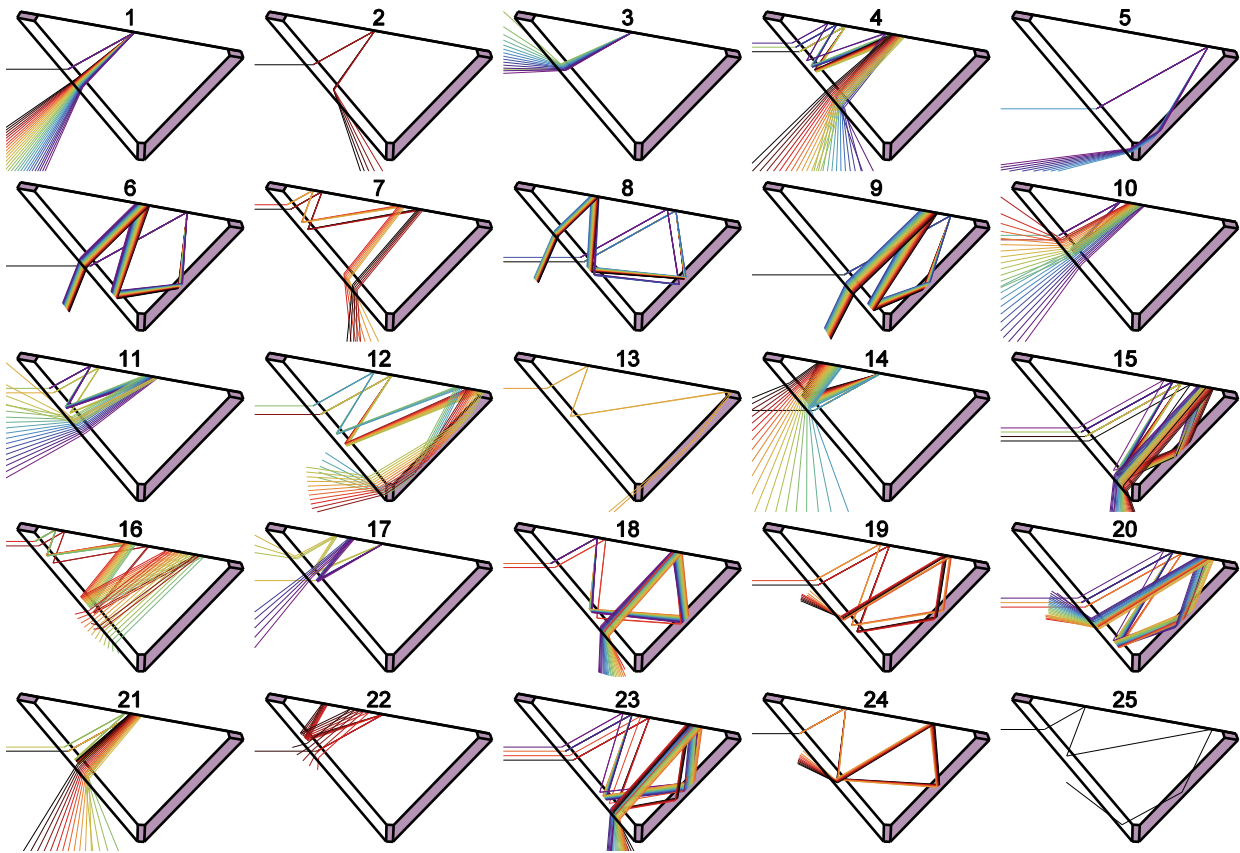

**Figure 3: Geometric plots of 25 different spurious internal reflection modes of the Sentinel-5 SWIR-1 BBM prism.**

Stray light having a spurious reflection on the third surface is minimized by adjusting the tilt of this surface and applying an absorbing coating. The entrance surface is tilted to ensure ghost reflections are directed out of the dispersed beam in the azimuthal direction.

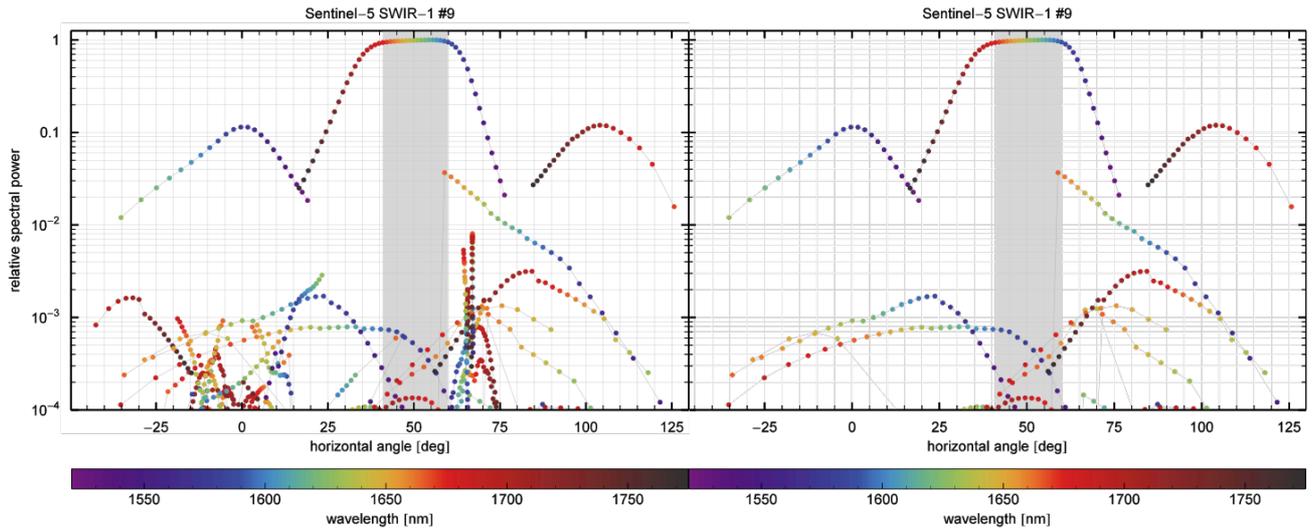

**Figure 4:** Results from ray-tracing simulations of the Sentinel-5 SWIR-1 BBM immersed grating. The plots show the relative spectral power as a function of horizontal angle for the individual reflection modes and are colour-coded according to wavelength. The grey band shows is the angular range used for the dispersed beam. The left-hand plot shows all reflection modes. In the right-hand plot, those reflecting on the third grating surface have been omitted.

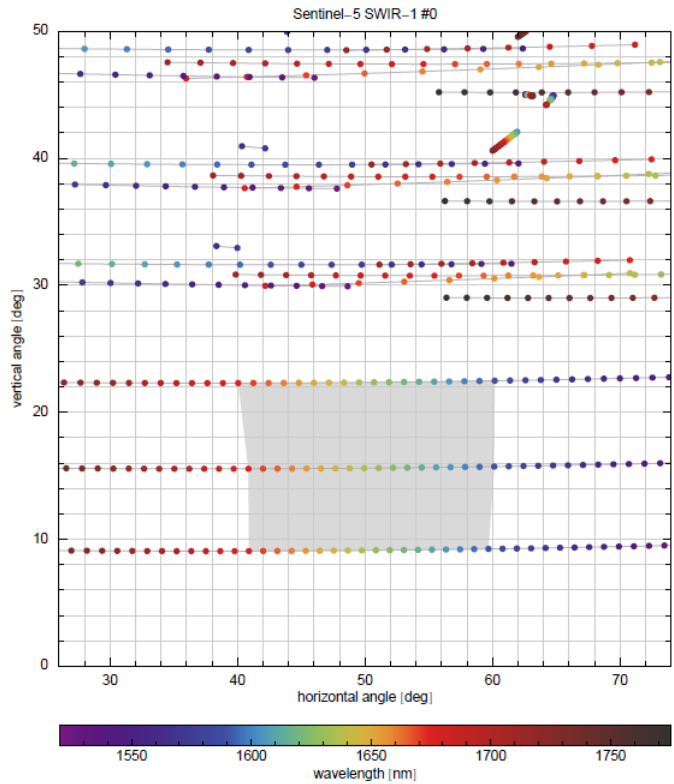

**Figure 5:** Results from ray tracing simulation of the Sentinel-5 SWIR-1 BBM prism showing the distribution of the main dispersed beam and spurious reflected light in the azimuthal plane. The grey area is the aperture of the downstream optics.

From the ray-tracing simulations the relative spectral power in the different modes can be plotted as a function of the dispersion angle and wavelength. This is shown in Figure 4. The effect of the optimization of the entrance surface tilt is seen in the azimuthal plane as presented in Figure 5. The final result of the ray-tracing optimization is a design in which the residual flux from spurious internal reflection in the direction of the dispersed beam is below the $10^{-4}$ level.

### 3. FABRICATION OF SILICON IMMERSED GRATINGS AT SRON

With the completion of the Sentinel-5 SWIR-1 BBM immersed grating, our current production method has been demonstrated at TRL 5. This production method, whereby the grating wafer and silicon prism are fabricated separately and then bonded together, can be split in 4 parts:

**Silicon prism fabrication**: Both the pre-cut monocrystalline silicon prism(s) and their polishing according to the design specification are sourced externally.

**Grating wafer fabrication**: A batch of silicon wafers with the required Total Thickness Variation (TTV) and off-cut angle (see Figure 2) is sourced from a semiconductor industry supplier. We currently use 150 mm wafers. The flatness and surface roughness of the wafer strongly influence the bond quality. We have found that post-polishing of the wafer after it has been patterned to decrease the microroughness and to remove thickness variation at mid-spatial frequencies ('orange-peel' effect) helps ensure a successful bond.

We start the lithographical process by determining the crystal lattice orientation using the method of Vangbo and Bäcklund[6]. The photolithographic patterning is done using an externally sourced mask, after which the etching of the grating grooves is performed.

**Bonding**: Before the bonding, the prism and wafer are cleaned and dried thoroughly as any contaminants on the bonding plane will result in local bonding defects. We use a Piranha solution followed by a RCA 'Standard Clean 1' process and a Marangoni-type dryer[7] for this process. Prior to bonding, the surfaces are activated with Reactive Ion Etching (RIE). Prism and wafer are then placed together in a custom bonder supplied by AML[8]. The bonding apparatus applies the wafer to the prism, creating the molecular bonds between the two surfaces.

After the bonding, an inspection using infrared imaging takes place to verify that the bonding is defect free. At this point, the wafer can still be removed from the prism relatively easily. After passing this inspection, the immersed grating is fused in an oven, strengthening the bond to approach the strength of bulk silicon. A second IR inspection is performed after the fusing. Finally, the excess wafer is trimmed.

**Coating application**: After successful bonding, optical coatings are applied to the relevant surfaces of the immersed grating. We have so far applied the reflective coating on the grating surface ourselves. For the antireflection and absorbing coating of the entrance and third surface, the immersed grating is sent to thin-film coating specialist CILAS (France).

With the coatings applied, the immersed grating is optically complete. We typically perform the performance characterization tests, as described in Section 5, before the grating is integrated with its mount. For space qualification, environmental tests, such as thermal vacuum and vibration testing, are performed after this integration, after which the performance tests may be repeated.

A key advantage of this method of immersed grating fabrication is the risk reduction achieved by splitting the grating wafer manufacture and the prism polishing. Several polished prisms may be procured and a batch of grating wafers produced. After inspection/testing of the wafers a ranking is made and a prime and backup wafer assigned to each prism. A bonding test can be performed with the prism with the least favourable polishing result and one of the lower-ranked grating wafers. Then the bonding of the prime prism/wafer combination can be performed. The inspection and characterization of the grating wafers is discussed in the next section.

## 4. INSPECTION AND CHARACTERIZATION OF GRATING WAFERS

The inspection and characterization of the grating wafers is performed with the aim of verifying the quality of the several aspects of the lithographic process. First, an inspection of defects is performed using an inspection tool to map the defects followed by detailed inspections with a Scanning Electron Microscope (SEM). Sample results from such an inspection are presented in Figure 6. Grating wafers are selected based on requirements on both the total number of defects and the maximum size of individual defects. We typically achieve a defect density on the order of $10^{-5}$ (fractional surface affected by defects).

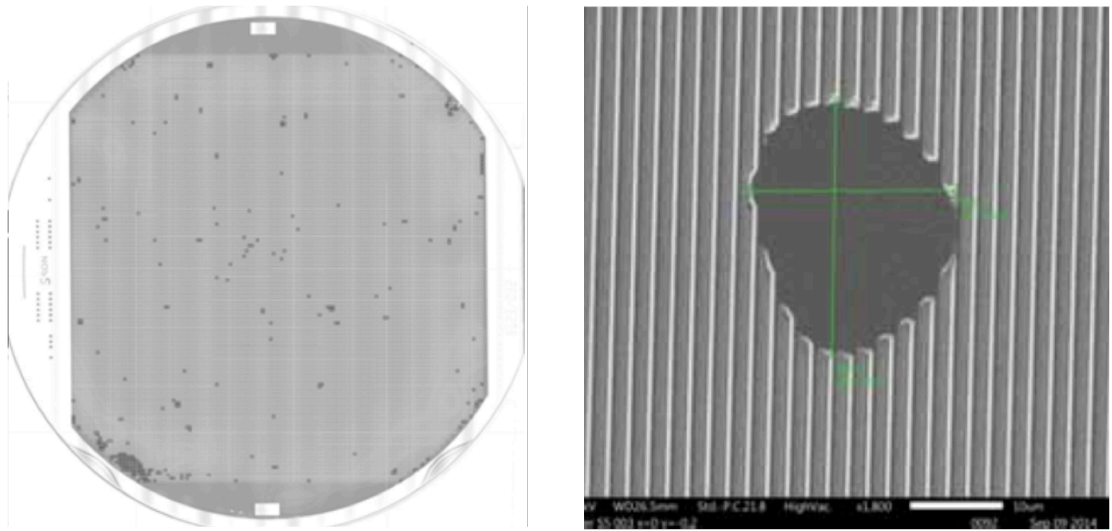

**Figure 6: Mapping of defects over a finished grating wafer (left) and an image of a sample defect (right). The images have been made using a Scanning Electron Microscope (SEM).**

Grating wafers that pass this inspection are subjected to a characterization of the uniformity of the etched groove pattern. A statistically meaningful sample of the groove periods and dam widths across the grating pattern is taken. This is done using tools such as an Atomic Force Microscope (AFM) or a SEM. We set requirements on both the deviation of the mean value of these properties from the targeted value and the standard deviation. A sample of the results from such a characterization is presented in Figure 7. We require a uniformity of dam widths over the grating clear aperture below 50 nm rms. Typically 25 nm rms is achieved.

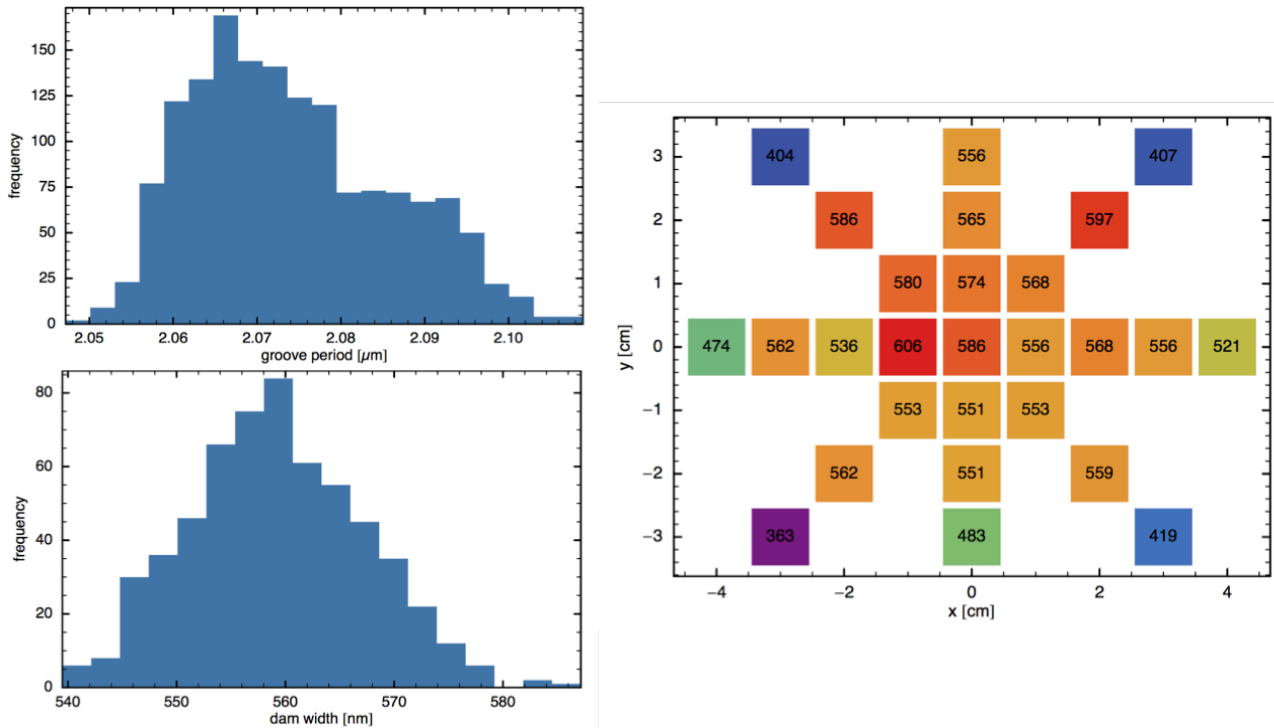

**Figure 7: Results from Atomic Force Microscope (AFM) and Scanning Electron Microscope (SEM) inspections of grating groove periods and dam widths of a grating wafer for the Sentinel-5 SWIR-1 BBM immersed grating. Two plots at left give the distributions of the period and dam widths obtained with the AFM. The plot at right shows the averaged dam widths for different locations on the grating wafer as measured with the SEM.**

In the results Figure 7 the dam width uniformity has been verified with both methods (AFM and SEM microscopy). The results are in agreement, increasing the confidence in the achieved grating quality.

Even if the specified mean and standard deviation of the error distribution has been achieved, a non-random spatial distribution of the errors over the grating surface may still cause unwanted results. In particular, a slow drift of the grating period or a slight curvature of the grooves will cause an increase in the WFE. For this reason, mapping of the errors over the grating surface, as shown in the right-hand part of Figure 7, is an essential verification step. In one case we have used a measuring microscope, where the grating wafer is put on a translating table of which the position can be determined with nanometer precision using laser interferometry, to check the absolute position of the grooves over the full grating surface.

Another important characteristic of the grating pattern is the blaze angle. Due to the resolution of the SEM images, the angle can only be verified with an accuracy of about one degree. In principle, the blaze angle is determined by the off-cut angle of the silicon wafer, which can be specified to a tenth of a degree. We are currently developing a method to verify the blaze angle with the same accuracy.

## 5. CHARACTERIZING THE PERFORMANCE OF SILICON IMMERSED GRATINGS

Once the immersed grating has been completed, the verification of the three key performance parameters, grating efficiency, wavefront error (WFE) and scattered stray light, takes place:

**Grating efficiency**

We measure the grating efficiency using a modified Perkin-Elmer Lambda 950 photospectrometer. The results for the Sentinel-5 SWIR-1 BBM immersed grating are presented in Figure 8.

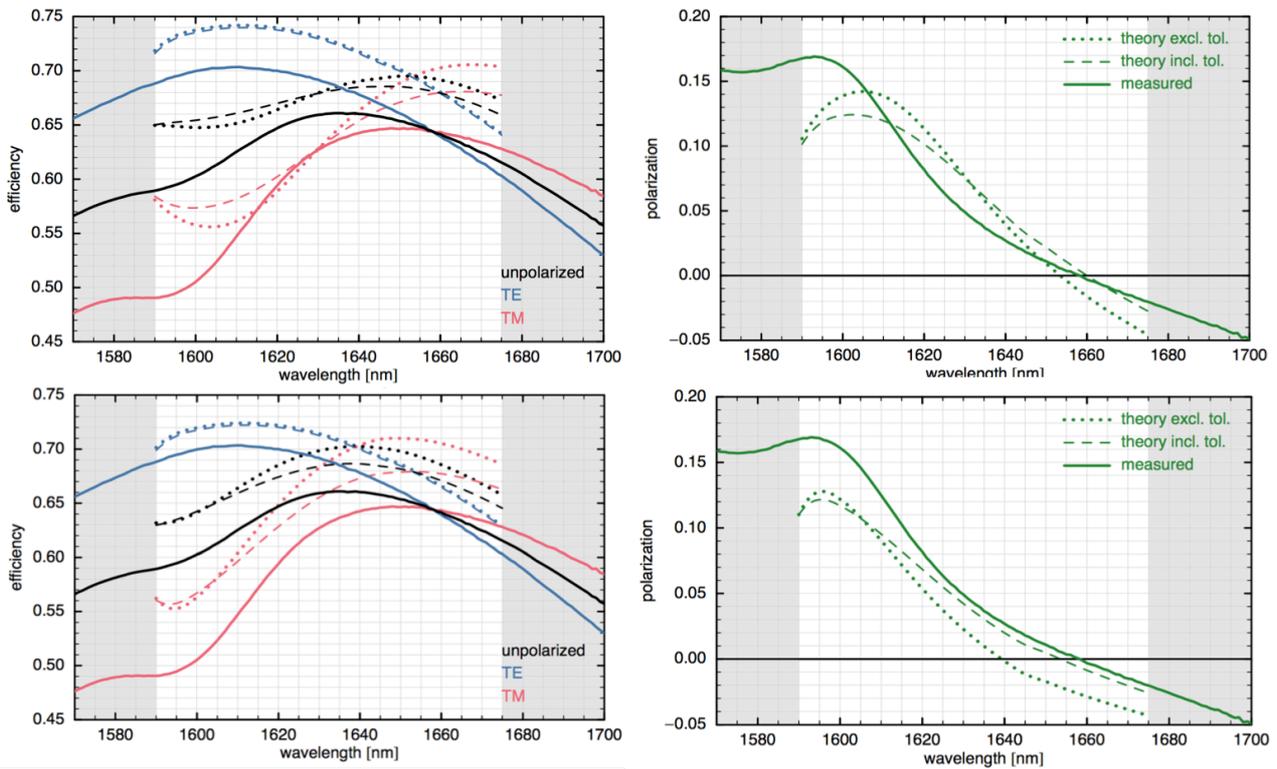

**Figure 8:** Plots of the measured grating efficiency compared with simulations. The two plots on the left side provide the total measured efficiency (black) and split into the TE and TM polarizations (blue, red) compared to simulations excluding (dotted) and including (dashed) tolerances. The two plots at right provide the polarization efficiency differential, being the difference in efficiency of the TE and TM states. In the top two plots, the simulations use the average measured dam widths, whereas in the bottom row, the dam width used for the simulation has been increased by 50 nm.

Over the operational wavelength range, the average (unpolarized) efficiency is 0.59 and the average efficiency differential between orthogonal polarization states is 0.17, meeting the requirements (0.59 and 0.20). We see however that the efficiencies are about 5% lower than what was expected from simulations (Figure 8, top left). The reason has not yet been fully determined. A better match is obtained when increasing the dam width in the simulation by 50 nm (Figure 8, bottom left). However, as determined during the grating wafer inspection (Figure 7), the dam widths are well within spec. The origin of the lower measured efficiencies may also be in the measurement setup. In any case we expect that for the Sentinel-5 flight programme, now underway, the dam width can be optimized further, resulting in a higher efficiency.

**Induced wavefront aberration**

Ideally, the induced WFE of the immersed grating is measured in immersion in the infrared at the operational wavelengths. Due to the limited availability of infrared interferometers, we have taken a different approach on the Sentinel-5 SWIR-1 BBM immersed grating. We measure the combination of WFE introduced through the wafer TTV and the groove pattern with a visible light interferometer (Zygo Verifire) from the outside and combine the results with the interferometry measurements of the polishing of the prism (entrance and grating) surfaces.

For the Sentinel-5 SWIR-1 BBM immersed grating, the interferometry image of the grating surface induced optical path difference is presented in Figure 9. The ellipsoidal shape in the interferogram may well be explained by a structured TTV from the wafer polishing.

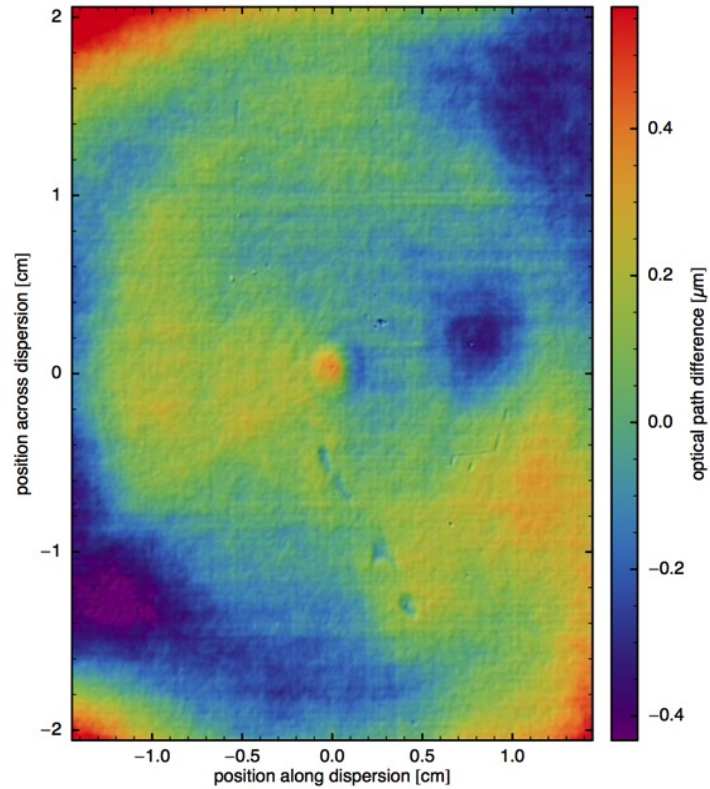

**Figure 9: Image of the induced WFE of the grating surface of the Sentinel-5 SWIR-1 BBM immersed grating, expressed in terms of optical path difference. The measurement was performed from the outside with an interferometer at visible wavelengths in Littrow configuration (order 4).**

The total WFE of the immersed grating can be obtained by combining these results with the residual prism surface shape errors after polishing according to those from the prism polishing according tot Eq. 1. The requirement on the WFE of the immersed grating is typically expressed as residual WFE after subtraction of piston, tip, tilt (Zernike modes $Z_0$, $Z_1$, $Z_2$), focus and (horizontal/vertical) astigmatism ($Z_3$ and $Z_4$). This last correction is also referred to as the 'debisphered' WFE.

In decomposing the WFE in Zernike modes we use the Zernike polynomial formalism derived for rectangular apertures by Mahajan et al.[9]. Using the classical decomposition for a circular aperture would result in an incorrect representation of the WFE as either a substantial part of the surface would not be taken into account (inscribed circle) or a large part of the surface would artificially be taken as flat (external circle).

For the Sentinel-5 SWIR-1 BBM immersed grating, the achieved 'debisphered' WFE is 153 nm, meeting (almost) the requirement of 150 nm.

**Straylight performance**

The straylight performance is verified using a Complete Angle Scatter Instrument (CASI) setup[10] at the European Space Agency's ESTEC establishment. In this setup, a sample can be placed at different incidence angles of a (tunable) laser source. A detector on a rotating arm then measures the azimuthal illumination profile. By rotating the sample 90° a second profile can be measured, which, combined with the first, results in the Bi-directional Reflectance Distribution Function (BRDF). The results for the Sentinel-5 SWIR-1 BBM immersed grating are presented in Figure 10.

In the figure, we compare the measurement with simulated profiles of mirrors with different surface roughness according to an adaptation of the Wein model[11, 12, 13]. The calculated mirror BRDF is given by Equation 2 where $\theta_{in}$ is the input and $\theta_s$ the scattering angle, $\sigma$ the surface roughness (rms height) and $k$ the wavenumber $k = 2\pi/\lambda$.

$$BRDF\left(\theta_{s}\right)=\frac{\Delta n}{\pi}\frac{k^{4}\sigma^{2}l^{2}}{1+[kl(sin\theta_{s}-sin\theta_{in})]^{2}} \tag{2}$$

The grating profile corresponds to the profile of a mirror with a surface roughness between 2 and 10 nm. This is compatible within the acquired accuracy with the roughness of 3 nm as specified in the requirements. Note that the figure also includes the measured profile of a reference mirror with 0.5 nm roughness as measured with white light interferometry. The reference mirror profile is above the theoretical curve for 1 nm, confirming the need for a more elaborate theoretical description. We are currently working on this topic and results will be presented elsewhere.

Note that in Figure 10, the peak of the reflectance curve at 0° is broader than that of the laser signature and the measured reference mirror. The broadness of this peak is a measure of how well focused the CASI setup was during the measurement, and does not influence the resulting straylight profile in the wings. In earlier test measurements of the immersed grating in the CASI setup, the sharpness of the peak approached that of the signature measurement.

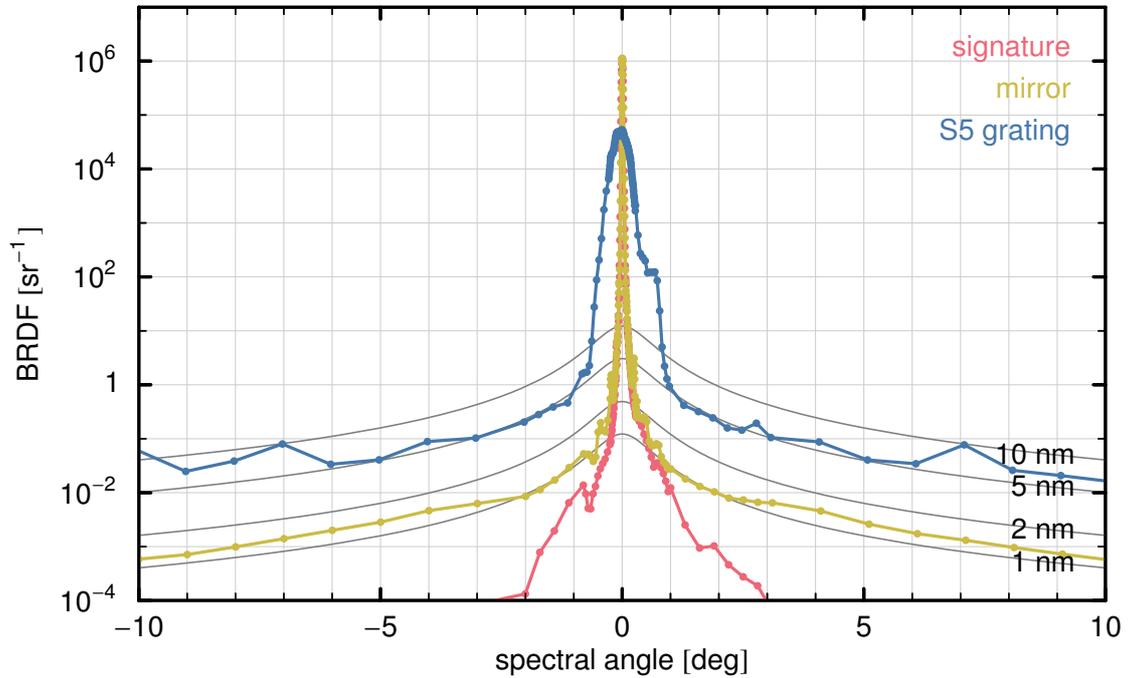

**Figure 10: Azimuthal (in the plane of dispersion) BRDF straylight profile of the Sentinel-5 SWIR-1 BBM immersed grating. The laser profile (measurement without a sample) is in red. The yellow curve is the profile of a mirror with a (stated) surface roughness of 0.5 nm. The grating curve is in blue. In grey are theoretical profiles for mirrors with different surface roughness.**

## 6. CONCLUSIONS

In this paper, we have detailed the design tools and verification methods employed at SRON Netherlands Institute for Space Research for optimizing three key performance parameters of silicon immersed gratings: Efficiency, wavefront error and straylight performance. With the qualification of the Sentinel-5 SWIR-1 BBM immersed grating, the method of assembling immersed gratings from separately fabricated monocrystalline prisms and grating wafers has reached a

Technological Readiness Level of 5. We have now embarked on the development and fabrication of the immersed gratings (SWIR-1 and SWIR-3) for the Sentinel-5 flight programme.

## 7. ACKNOWLEDGEMENTS

The development of the Sentinel-5 SWIR-1 BBM immersed grating was performed in close collaboration with Airbus Defence & Space. We thank Ramon Vink and Volker Kirschner for taking the BRDF measurements at ESTEC. We thank Hélène Krol at CILAS for application of the optical coatings. Finally, we thank the SRON project team for their contribution to the presented work.